\documentclass{optica-article}

\journal{opticajournal} 

\articletype{Research Article}

\usepackage{lineno}
\usepackage{soul}
\usepackage{graphicx}
\usepackage{amsmath}
\usepackage{color}
\usepackage{hyperref}
\usepackage{siunitx}
\usepackage{soul}
\usepackage{gensymb}
\usepackage{textcomp}
\usepackage{acronym}
\usepackage{caption}
\usepackage{mhchem}

\newcommand{\TrtHz}{$\mathrm{T}/\sqrt{\mathrm{Hz}}$}

\begin{document}


\newacro{WGM}{Whispering Gallery Mode}
\newacro{EBPG}{Electron-beam Pattern Generator}
\newacro{EBL}{Electron-beam Lithography}
\newacro{RIE}{Reactive-ion Etching}
\newacro{IPA}{Isopropyl Alcohol}
\newacro{FSR}{Free Spectral Range}
\newacro{SNR}{Signal-to-Noise Ratio}
\newacro{SEM}{Scanning Electron Microscope}
\newacro{FEM}{Finite Element Method}
\newacro{Si}{silicon}
\newacro{SiO$_2$}{silica}
\newacro{DFB}{Distributed Feedback}
\newacro{PD}{photodector}
\newacro{PSD}{Power Spectrum Density}
\newacro{FWHM}{Full Width at Half Maximum}

\title{Waveguide-integrated and portable optomechanical magnetometer}

\author{Fernando Gotardo\authormark{1,2}, Benjamin J. Carey\authormark{1,2}, Hamish Greenall\authormark{1,2}, Glen I. Harris\authormark{1,2},  Erick Romero\authormark{1,2}, Douglas Bulla\authormark{4}, Elizabeth M. Bridge\authormark{5}, James S. Bennett\authormark{3}, Scott Foster\authormark{4}, and Warwick P. Bowen\authormark{1,2*}}

\address{\authormark{1}School of Mathematics and Physics, The University of Queensland, St Lucia, Queensland 4067, Australia.\\
\authormark{2}ARC Centre of Excellence for Engineered Quantum Systems, St Lucia, Queensland 4067, Australia.\\
\authormark{3}Centre for Quantum Dynamics, Griffith University, Nathan, Queensland 4072, Australia.\\
\authormark{4}Australian Government Department of Defence Science and Technology, Edinburgh,\\ South Australia 5111, Australia.\\
\authormark{5}Quantum Brilliance, Acton, Australian Capital Territory 2601, Australia}
\email{\authormark{*}w.bowen@uq.edu.au} 


\begin{abstract*} 
Optomechanical magnetometers enable highly sensitive magnetic field sensing. However, all such magnetometers to date have been optically excited and read-out either via free space or a tapered optical fiber. This limits their scalability and integrability, and ultimately their range of applications. Here, we present an optomechanical magnetometer that is excited and read-out via a suspended optical waveguide fabricated on the same silicon chip as the magnetometer. Moreover, we demonstrate that thermomechanical noise limited sensitivity is possible using portable electronics and laser. The magnetometer employs a silica microdisk resonator selectively sputtered with a magnetostrictive film of galfenol (FeGa) which induces a resonant frequency shift in response to an external magnetic field. 
Experimental results reveal the retention of high quality-factor optical whispering gallery mode resonances whilst also demonstrating high sensitivity and dynamic range in ambient conditions. The use of off-the-shelf portable electronics without compromising sensor performance demonstrates promise for applications.
\end{abstract*}



\section{Introduction}
%
In recent years, optomechanical sensors have emerged as a powerful new type of sensor for stimuli ranging from temperature~\cite{Purdy2017} to pressure~\cite{Basiri2019}, forces~\cite{Gavartin2012, Harris2013} and acceleration~\cite{Krause2012}. Such sensors leverage both optical and mechanical resonances to enable high sensitivity and high spatial resolution. Optomechanical magnetometers are one example~\cite{Forstner2012,Forstner2014,Li:20,yu2016optomechanical}, that are attractive due to the crucial role that  
high-sensitivity magnetometers play  in applications 
ranging from fundamental research to medical diagnostics~\cite{Savukov2013, glenn2018high}, mineral exploration and surveying \cite{Meyer2005, Edelstein_2007}, magnetic anomaly detection~\cite{li2015detection, Sheinker2009}, and navigation~\cite{Bennett2021, budker2007optical, rodriguez2019science}.  
Owing to their photonic nature, they are light-weight, small-sized, low power~\cite{Bennett2021,Forstner2014,Li:18Quant}, and  can be exceedingly resilient to detriments such as electrical interference and radiation. 

At their current state of development, optomechanical magnetometers 
have achieved tens-of-micron spatial resolution with sensitivities ranging from several n\TrtHz{} down to tens of p\TrtHz{} \cite{Li2018,Forstner2014, Bennett2021, Li:20}. The demonstrated sensitivity is competitive with SQUID and diamond magnetometers of similar size but without the need for cryogenics or high-powered optical and RF components \cite{Bennett2021, zhu2017polymer, Yu2018}, with theoretical models suggesting that sensitivities in the low, or even sub-, femtotesla may be possible in future \cite{bowen2017cavity}.

To-date, optomechanical magnetometers have used free-space or tapered-fiber coupling for optical excitation and readout~\cite{Yu2018, romero2018quantum, yu2016optomechanical}. This prevents them from being fully integrated on a silicon chip. Furthermore, those demonstrations used the magnetostrictive material Terfenol-D to convert magnetic fields into a measurable strain~\cite{Forstner2012,Forstner2014}. Which is difficult to reproducibly deposit and sensitive to corrosion and oxidation~\cite{Nivedita2017galf}. Works with this material have also relied on high performance laser and electronic systems~\cite{Li:18Quant}. Together, this introduces significant challenges for applications outside of laboratory environments. The work reported here seeks to address these challenges.


We develop an optomechanical magnetometer that is efficiently coupled to an on-chip suspended waveguide,
by employing  galfenol (Fe$_{82}$Ga$_{18}$) for the first time to convert magnetic fields to a measurable mechanical signal. 
This provides low-cost sputter-coated thin-films with improved resilience to corrosion and oxidation \cite{Nivedita2017galf} and good magnetostriction ($\sim$400~ppm) at lower saturation fields\cite{Clark2005galf,Nivedita2017galf}. We also use portable electronic and laser systems to control and read-out the magnetometer, showing that they allow performance that is limited by fundamental thermomechanical noise rather than laser or electronic noise. Together, this represents progress towards robust, portable and high performance mangetometers that could be employed in diverse research and industrial settings.

\section{Design and Simulation}
\label{sec:design}
\subsection{Device Design and Functionality}
\begin{figure}[h]
 \centering
    \includegraphics[width=0.7\textwidth]{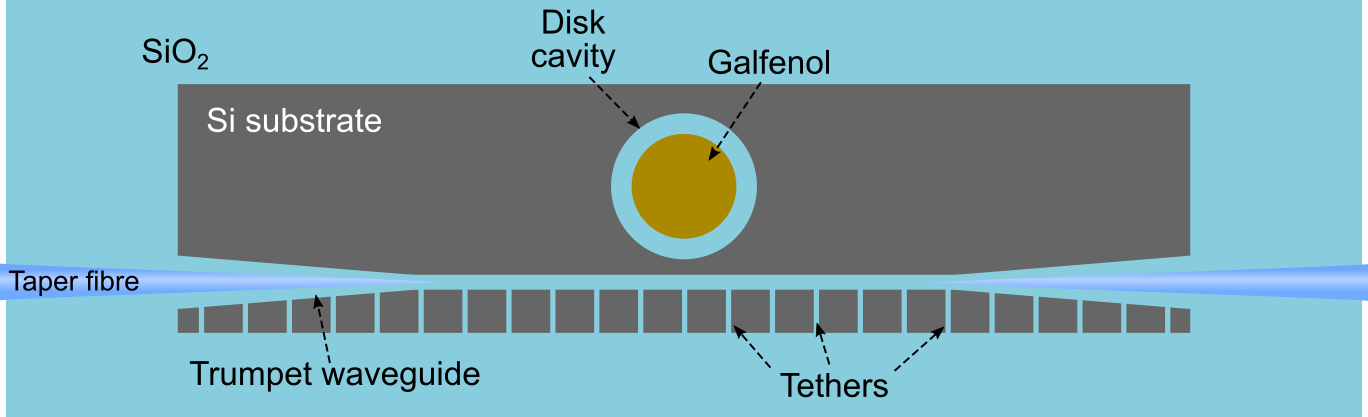}

    \caption{Design of the integrated SiO\textsubscript{2} magnetometer. Here, laser light is coupled to the trumpet waveguide via a tapered fiber. The waveguide is narrow at the centre to optimize evanescent field coupling to the disk cavity. The galfenol is sputtered on top of the cavity. The tethers support the waveguide, preventing buckling.}
    \label{fig:SiO2 mag}
\end{figure}

The device design concept is depicted in Figure \ref{fig:SiO2 mag}. It is based around a 100~$\upmu$m-diameter silica microdisk cavity on a silicon chip. This microdisk is capable of supporting optical whispering galley modes (WGMs) throughout the visible and near-infrared spectrum as well as megahertz frequency mechanical resonances.
A 1.5~$\upmu$m wide silica waveguide is fabricated from the same layer of silica as the microdisk. Both microdisk and waveguide are undercut so that the optical modes are confined to the silica, rather than leaking into the higher-refractive index silicon substrate. The microdisk is suspended from a central silicon pedestal. The waveguide is suspended using thin silica tethers that are patterned along its length to reduce warping of the waveguide that can be caused by the intrinsic stress present in the as-fabricated SiO\textsubscript{2} films. Buckling of the waveguide could lead to severe bending losses, and out-of-plane buckling can lead to inconsistent coupling between the waveguide and optical cavity.
The tethers are sub-wavelength (240~nm width) in order to minimise optical scattering of light propagating within the waveguide. As silica is lower refractive index than many other waveguide material (\textit{e.g.}, silicon) the guided wavelength is longer allowing for minimal scattering. The waveguide is broadened (inverse-tapered) at input and output to efficiently mode-match light into and out-of tapered optical fibers. 


The microdisk is coated with galfenol in a disk of diameter sufficiently smaller than the disk diameter so as not to introduce optical absorption. Galfenol is chosen because of its high magnetostriction, low fragility, and low volatility~\cite{Nivedita2017galf}.  
When a magnetic field is applied, the expansion of the galfenol induces strain in the microdisk, changing the optical path length and hence the optical resonance frequency. The magnetostrictive response is amplified at frequencies close to mechanical resonances of the microdisk,
leading to enhanced modulation of the intracavity optical field.

The light is coupled into the disk evanescently from the suspended waveguides which are designed to support single mode propagation around 1550~nm. The coupling of light from an optical fiber into the on-chip waveguide despite the geometric mismatch (SMF-28 optical fiber has core and cladding diameters of 10~$\upmu$m compared to the 300~nm thickness of the waveguide) is facilitated by mode-matched taper-drawn fibers (similar to those presented in \cite{Ward2014pull}) and trumpeted waveguide (4~$\upmu$m down to 1.5~$\upmu$m over a length of 30~$\upmu$m). This allows adiabatic coupling of light to and from the waveguides. 
Further, the 4~$\upmu$m wide flared section acts as a semi-rigid anchor point for the fiber, and its size reduces the requirement for extremely precise positioning. This allows greater tolerance to imperfections in important integration processes, such as bonding the fiber tip in place. 

\subsubsection{Finite-element Simulations}

\begin{figure}
    \centering
    \includegraphics[width=0.95\textwidth]{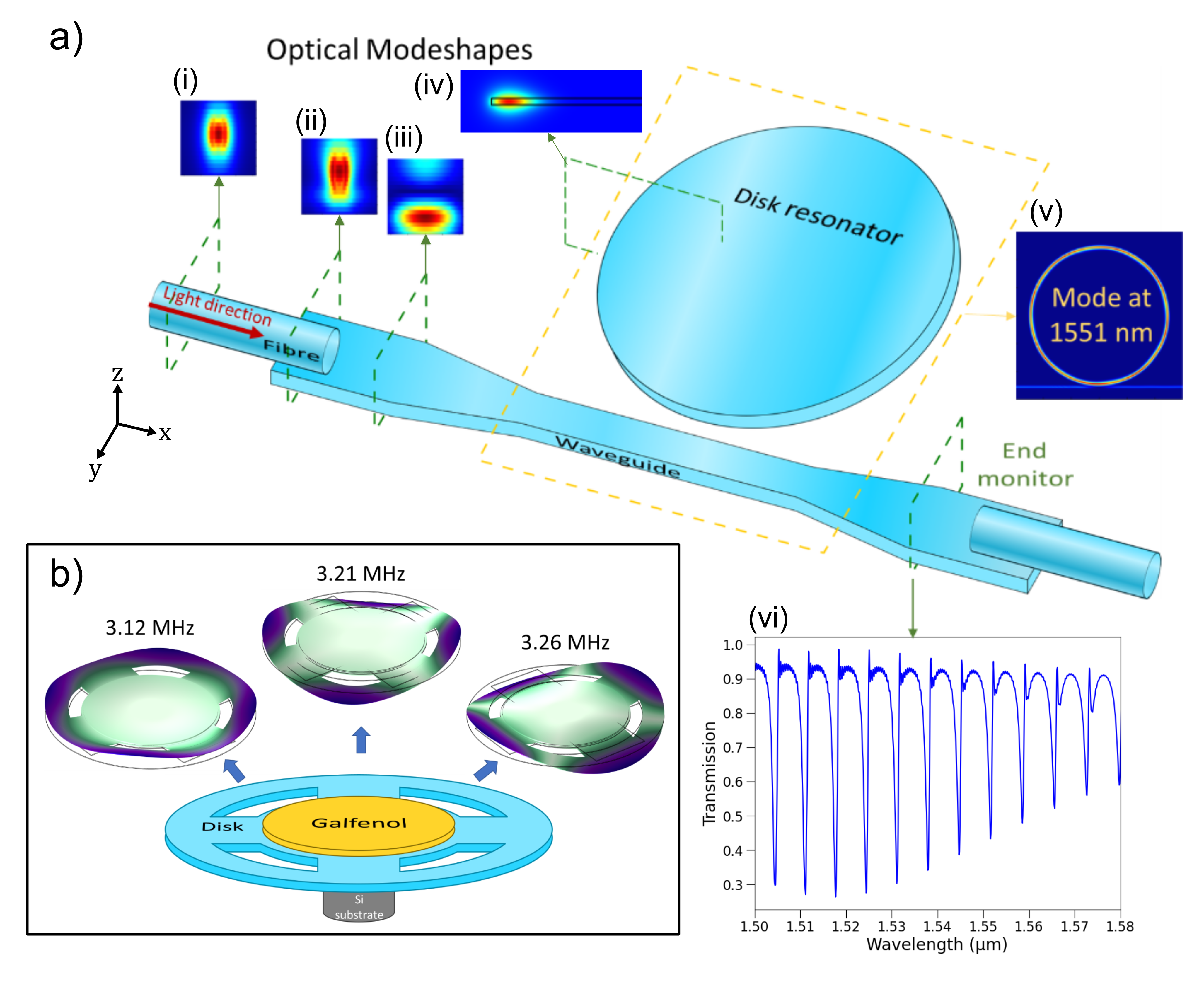}
    \caption{\ac{FEM} simulations of the integrated optomechanical magnetometer.\\
    a) The optical properties of the device. The cross-sectional optical mode profiles (i-iii) at their corresponding green rectangles demonstrate the evolution of the optical modes from fiber to SiO\textsubscript{2} trumpet waveguide and (iv) the cross-sectional optical mode inside the microdisk. (v) Depicts the optical mode at the planar cross-section (yellow rectangle) of the device, and (vi) shows the optical transmission spectrum of the system. (b) Mechanical simulation revealing the resonance frequencies and their flexual mode-shapes}
    \label{fig:fig2}
    \label{fig:SiO2 sim opt}
    
\end{figure}

\ac{FEM} simulations performed with Ansys-Lumerical software for the optical performance of the device are presented in Figure \ref{fig:fig2} a).  Here the coupling from fiber to waveguide is studied by monitoring the optical mode cross-section ($yz$) along the propagation direction ($x$). The cross-sections shown in (i-iii) correspond to cross-sections of only fiber (i), fiber \& waveguide (ii), and only waveguide (iii). For simplicity, both the fiber taper and the waveguide were considered uniform at the coupling region. The taper was chosen to have a 1~$\upmu$m diameter with the optimum waveguide width of 4~$\upmu$m then found using recursive simulations.
From the simulated cross-sections, we see that the optical mode migrates from the fiber into the waveguide. We obtain a fiber-to-waveguide coupling efficiency of 60\% by taking the ratio of the optical power contained within the fiber at point (i) and within the waveguide at point (iii).

Within the same simulation the waveguide-to-disk coupling and disk resonances were also studied. Here the optical excitation frequency was swept and the optical intensity across the geometry was recorded at each frequency.  The transmission efficiency across the device could then be calculated by comparing the integrated intensity over the cross-sections of the input and output of the waveguide (\textit{i.e.}, $T = I_{\text{out}}/I_{\text{in}}$, as measured at points (iii) \& (vi)). Fig. ~\ref{fig:fig2}(a)(vi) shows the expected periodic transmission dips when the frequency of the light matches WGMs of the microdisk. The transmission is predicted to drop by as much as 70\%, indicating that efficient coupling of light into WGMs should be possible. The correspondence of the dips with WGMs is confirmed in Fig.~\ref{fig:fig2} a)(v), which shows confinement of the light to the periphery of the disk when driven on resonance (at a wavelength of 1551~nm in this case).
To further corroborate the confinement of the WGMs we performed an axisymmetric eigenmode-solver \ac{FEM} simulation in COMSOL Multiphysics (Figure \ref{fig:fig2} a)~(iv)). This confirmed that the WGM is contained within the outer 5~$\upmu$m of the disk, as expected.

The \ac{FSR} of the optical resonances and corresponding coupling were calculated from the simulated transmission  in Fig. ~\ref{fig:fig2}(a)(vi). We find a simulated \ac{FSR} of approximately 7~nm. This compares well to the expected FSR given the circumference of the microdisk of:
\begin{equation}
    \Delta \lambda \approx \frac{\lambda^2}{n(\lambda)L} \approx 7.6 \;\mathrm{nm},
\end{equation}
where $n(\lambda)$ is the effective refractive index of the cavity mode (taken from Lumerical simulations to be 1.01) and $L$ is the length of the cavity \textit{i.e.}, $L =$ 100$\pi\;\mathrm{\upmu m}$.\\


As evidenced by the results in Figure \ref{fig:fig2} a)~(iv \& v), the optical field of the \ac{WGM} extends negligibly into the centre of the SiO\textsubscript{2} disk. Hence, the addition of the optically absorbing magnetostrictive layer to the disk's centre should not significantly affect the quality of the optical modes contained therein \cite{Li2018}.


Using COMSOL Multiphysics, we performed further simulations to assess the mechanical properties of the microdisk. As shown in Figure \ref{fig:fig2} b), we found the mechanical eigenmodes by using fully three-dimensional geometry of the released devices. This was necessitated because of the inclusion of stress-release slots (discussed in \ref{Sec:Fab}) that break the axial symmetry of the mechanical modes. The physical properties of the galfenol were taken from the datasheet supplied by TdVib \textit{LLC}.  
The lowest frequency mechanical flexural mode and two lowest frequency crown modes are shown in Figure \ref{fig:fig2}~b), with mechanical frequencies of 3.12, 3.21, and 3.26~MHz, respectively. 

\section{Device Fabrication}
\label{Sec:Fab}

\begin{figure}
    \centering
    \includegraphics[width=0.95\textwidth]{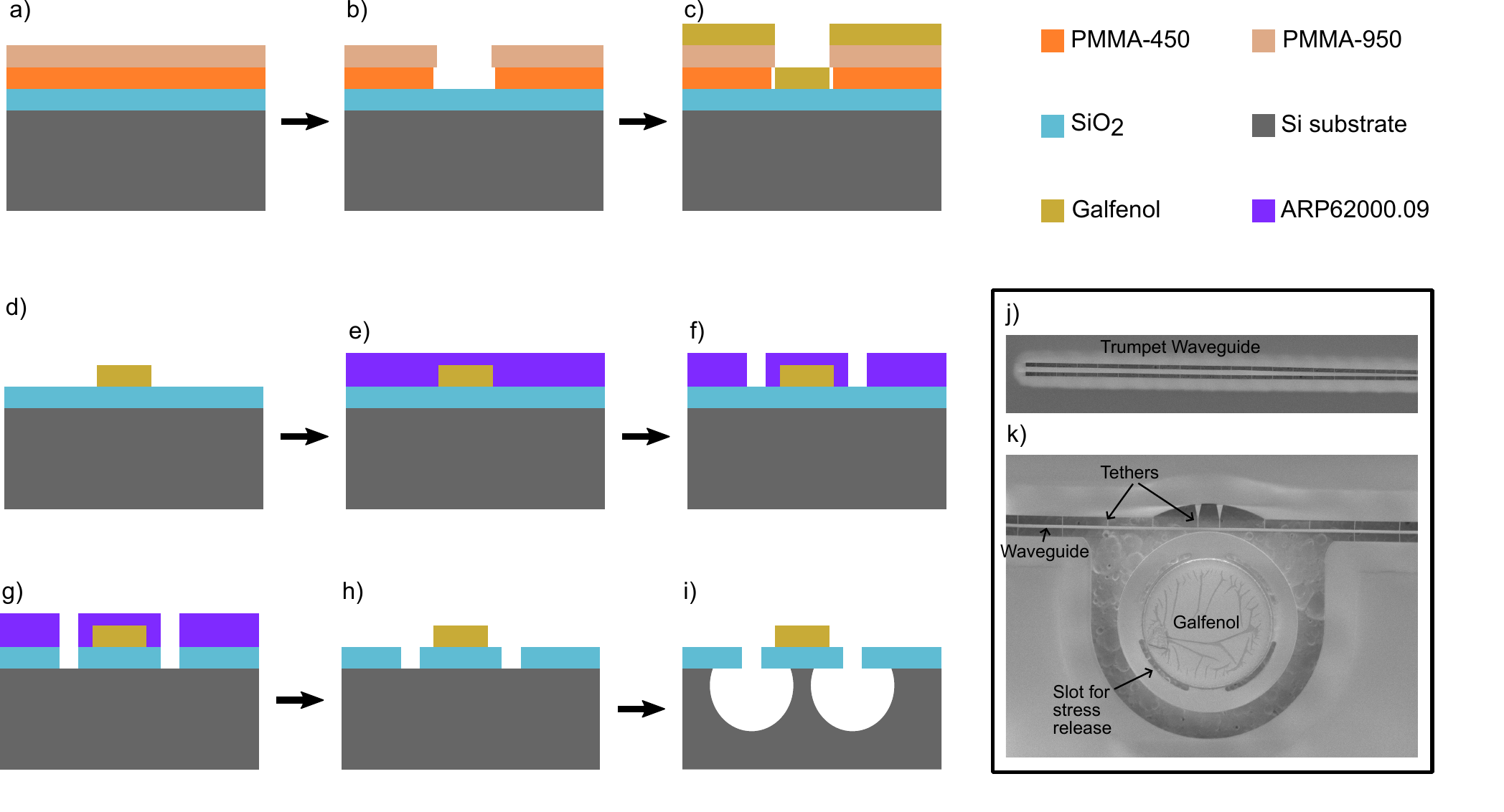}

    \caption{Representation of the fabrication process. a) PMMA e-beam resist deposition. b) \ac{EBPG} exposure and development. c). Galfenol sputter deposition. d) lift-off. e) ARP e-beam resist deposition. f) Second \ac{EBPG} exposure and development. g) \ac{RIE} of SiO\textsubscript{2}. h) Resist removal. i) Released devices after undercutting by XeF\textsubscript{2} etching of the Si layer. \\
    \ac{SEM} images of the final devices depicting j) trumpet wave guide and k) the optomechanical cavity with galfenol layer in its centre and coupling waveguide.}
    \label{fig:SiO2 SEMs}
    \label{fig:SiO2 mag recipe}
    \label{fig:fig3}
\end{figure}

The fabrication process used to produce the devices is outlined in Fig.~\ref{fig:fig3} a)~(i). SiO\textsubscript{2} (300~nm) on Si substrate wafers (500~$\upmu$m, 4") were 
diced into square 15$\times$15~mm chips, large enough to fit more than 100 devices per chip. 
Electron beam lithography was used to define patterns for galfenol deposition and markers for subsequent lithography steps
in the following way.
%
Two layers of PMMA resist were spin-coated (PMMA 950k and 495k at 2500 RPM) onto of the SiO\textsubscript{2}/Si substrate, then patterned with an \ac{EBPG} (Raith~EBPG~5150) with 100~kV accelerating voltage and a 1200~$\upmu$C/cm\textsuperscript{2} dosage. Post exposure, the chips were developed in methyl isobutyl ketone (MIBK) and rinsed with \ac{IPA}.

To produce the markers, 5~nm of Ti and 50~nm of Au were e-beam evaporator deposited (Temescal FC-2000) follow by a lift-process via submersion into acetone and \ac{IPA}.
The galfenol films were then sputtered by magnetron DC sputtering in an argon atmosphere (150~W, 2~mTorr) with a (3~inches~dia.) galfenol target. A seeding layer (5 nm Ta, plus a 16 nm Cu) and capping layer (5 nm Ta) were used to promote adhesion and inhibit corrosion respectively. Afterwards, the lift-off process was repeated, resulting in a 300~nm thick, 60~$\upmu$m diameter circular thin-film of galfenol on top of the SiO\textsubscript{2} layer.

With markers produced and galfenol deposited, the waveguide and the disk cavity structures were then defined. For this, 20~nm thick Ti prime adhesion helper was spin-coated (4000~rpm) and baked (150\degree{}C, 15 min) follow by a layer of ARP 6200.09 (CSAR-62, All Resist) 350~nm thick (1250~rpm spin-coat, 180\degree{}C for 5 min bake).  
The chip was then patterned with the Raith EBPG 5150 (100 kV, 260 $\upmu$C/cm\textsubscript{2}). Proximity effect correction was performed using GenISys Beamer software to ensure precision and reproducibility in the \ac{EBPG} process. Post exposure, the patterns were developed with All Resist AR600-546 and rinsed with o-xylene and \ac{IPA}.

\ac{RIE} was used remove the unwanted SiO\textsubscript{2} using an Oxford Instruments PlasmaPro 80. Here 25 sccm CHF\textsubscript{3} and 45 sccm Ar at 200 W RF power for 12 min anisotropically etched all the way through the SiO\textsubscript{2} layer exposing the silicon substrate. 50 sccm O\textsubscript{2} at 50 W was then used to remove any residual resist.
Finally, the SiO\textsubscript{2} structures were undercut by etching of the supporting silicon with Xenon Difluoride (XeF\textsubscript{2}) gas (SPTS XACTIX e2 Series XeF\textsubscript{2} etcher). Here, 10 pulses of XeF\textsubscript{2} gas at a pressure of 2 torr provides an isotropic etch rate of about 1.4~$\upmu$m per pulse with a selectivity of >1000:1. This removed the Si beneath the silica waveguide and WGMs of the microdisk whilst leaving both the silica layer and galfenol unmarred.

\ac{SEM} (Jeol 7800) imaging of the devices was performed in order to assess their structural integrity. Fig.~\ref{fig:fig3}~j) \& k) shows SEMs of the trumpet-shaped waveguide ends for fiber coupling and the waveguide near the resonator, supported by tethers to the main body of the wafer. It is apparent that the waveguide shows no signs of buckling or collapse after the release process. It can also be observed that the undercut beneath the silica layer is approximately 18 $\upmu$m. This undercut extends under the disk, leaving behind a \ac{Si} pedestal which is obscured by the galfenol coating above. Measurements on a device with no galfenol revealed a pedestal width of 15 $\upmu$m (measured with a Zeta 300 3D optical profiler).

Stress release slots in the resonator were found to be necessary to prevent buckling of the disk due to the inherent strain within both the SiO\textsubscript{2} layer and the galfenol film. However, as discussed in section \ref{sec:design}, these slots are expected to have negligible effect on the optical modes because they are outside of the region containing appreciable intensity. The mechanical simulation of Fig.~\ref{fig:fig2} b) fully accounts for the effect of the slots on the mechanical eigen-frequencies.

A critical parameter for consideration during fabrication of the device is the distance between the disk and the waveguide at the coupling region ($d$). As the light is coupled evanescently the coupling efficiency ($\kappa$) follows the relation $\kappa \propto e^{-d}$ \cite{Fang2017coupling}. Devices with a range of waveguide-microdisk coupling distances were fabricated  in order to produce resonators with optimum coupling strengths. The devices with near-critical coupling were further investigated.

\section{Device Performance}
\label{sec:perf}

\begin{figure}
   \centering
    \includegraphics[width=0.85\textwidth]{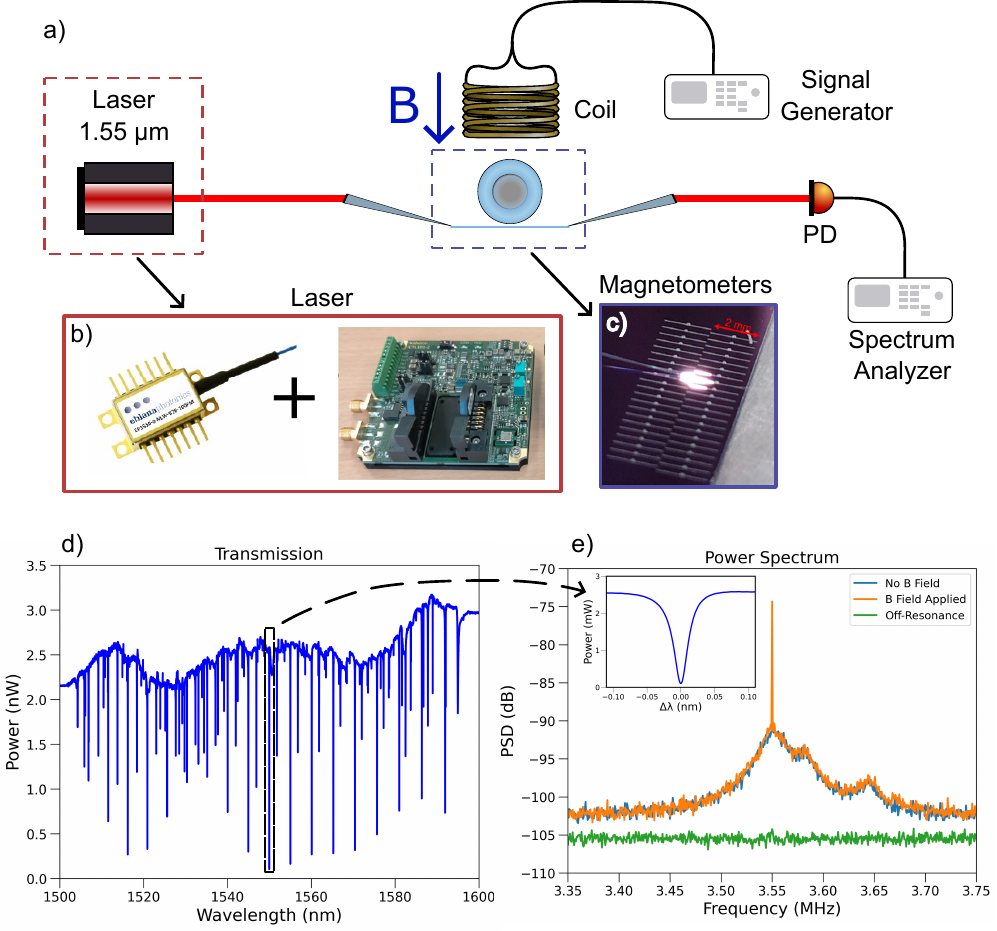}
    
    \caption{Experimental performance of the magnetometers. a) Schematic demonstrating the experimental setup with corresponding photographs of the laser system b), and devices on test-rig c) used for the measurements. d) Optical transmission spectra of the devices with accompanying high-resolution spectra around one of the \ac{WGM} resonances (inset of e) which was used for the sensing investigations. e) Power spectrum of the transmitted light with and without externally applied magnetic field.}
    \label{fig:SiO2 optical}
    \label{fig:SiO2 sens}
    \label{fig:fig4}
\end{figure}

The experimental setup used to assess the performance of the integrated magnetometers is depicted in Fig.~\ref{fig:fig4}a). Here, a continuously tuneable laser (EXFO T100S-HP) supplied light to the resonator via tapered optical fibers with a house-built test rig featuring two 3-axis translation stages for precise positioning of the fibers. The transmitted light was then guided to a Thorlabs (PDA100-CS) \ac{PD} and the photocurrent was analyzed with a spectrum analyzer (Siglent SSA 3021x). A function generator (Rigol DG-1022) was used to directly drive a home-wound coil (8~turns, 10~mm dia.) held approximately 1~mm above the chip, producing a field of $\sim$50~$\upmu$T$_{PP}$ at the surface of the chip (drive voltage of 10~V$_{PP}$). 

The emission wavelength of the laser was swept and the voltage output from the \ac{PD} (and hence power via the known responsivity and transimpedance gain of the \ac{PD}) was recorded to characterise the optical mode spectrum of fabricated devices.
The optical transmission spectrum of a typical device is presented in Fig.~\ref{fig:SiO2 optical}~d), showing many dips in the transmission dips each associated to one WGM. The observed \ac{FSR} of $\approx$ 7~nm is in good agreement with the \ac{FSR} as determined from the \ac{FEM} seen in Section \ref{sec:design} and Fig.~\ref{fig:fig2}a)(vi). On this device (with a designed waveguide-microdisk separation of 550~nm) we find that the WGM at a wavelength of 1551~nm (enclosed by the dashed box in Fig.~\ref{fig:SiO2 optical}~d)) is close to critically coupled, with a transmission dip of~$\sim95\%$.


Because the 1551~nm WMG mode is close to critically coupled, we select it to perform magnetic field measurements. A high-resolution sweep across the mode is shown in the inset of Fig. ~\ref{fig:fig4}e). From this the optical $Q$ of the cavity is estimated to be: 
\begin{equation}
    Q_{opt}\approx \frac{\lambda_0}{\Delta\lambda_{FWHM}} \approx10^5 .
\end{equation}

For many applications, it is desirable to use a low cost, low power, and compact laser source, together with compact electronic systems, rather than the high performance EXFO fiber-laser and associated electronics used in this work to date. Here, we test whether it is possible to do this without sacrificing performance. A commercially available \ac{DFB} laser (Eblana EP1550) with a portable laser driver (Koheron CTL101) was used to couple light onto and off the chip (Figure \ref{fig:fig4} b)) were used for all subsequent measurements.

Tuning the \ac{DFB} laser to the side of the 1551~nm WGM allows shifts in the resonance frequency to be directly observed as changes in the optical intensity. This allows optical detection of mechanical vibrations, and hence magnetic field, without the need 
for interferometric detection \cite{Forstner2012}. Analysing the resulting photocurrent on a spectrum analyser reveals the mechanical mode spectrum shown in Fig.~\ref{fig:fig4}b). Three mechanical modes are observed at frequencies of at 3.55, 3.58, and 3.64 MHz. We attribute the discrepancy between the measured and simulated mechanical frequencies to the inherent stress of the galfenol film ($\sigma$=500~MPa) adding a stiffening effect to the mechanical resonances.

The noise-floor of the measurement consists of two components. At frequencies far away from the mechanical resonance frequencies it is dominated by laser noise. This is evidenced by an increase in noise when the laser tuned to the side of the WGM compared to when it is at a frequency far away from the mode. At frequencies close to the mechanical resonance frequencies, it is dominated by the thermomechanical noise. We can therefore conclude that the compact electronic systems used introduce no degradation in performance and, close to the mechanical resonances, neither does the optical noise of the DFB laser.


To determine the magnetic field sensitivity of the device, we apply a magnetic field at the frequency of the most prevalent mechanical mode (3.55 MHz). This induces a sharp peak in the \ac{PSD} (Figure \ref{fig:fig4}e) orange-trace), evidencing that magnetic fields can be detected.
With this particular applied field (B\textsubscript{AC} = 50 $\upmu$T) we measure a \ac{SNR} of 17.5~dB. The magnetic sensitivity of the device at 3.55 MHz 
could then be calculated using:
\begin{equation}
    S = B_{AC} \left({10^{\frac{SNR}{10}}\times \mathrm{RBW}} \right)^{-1/2}
\end{equation}
where RBW is the resolution bandwidth of the spectrum analyser \cite{Forstner2012}. This yielded a sensitivity of 2~$\upmu$\TrtHz, which is comparatively less sensitive than previously demonstrated optomechanical magnetometers that present sub n\TrtHz\ sensitivities \cite{Bennett2021}.\\
This reduced sensitivity can be attributed to geometric design of the device. With these devices the galfenol lies in part, above the pedestal, where the silicon greatly suppresses both  mechanical motion and imbued strain. Further, the mechanical eigenmodes have very little motion where the galfenol resides, thus do not experience the maximum possible driving force from the magnetostriction. These effects provide a reduction of the force exerted onto the optical eigenmodes from magnetostrictive stress.
Thus, the sensitivity could be considerably improved through the use of device geometry optimized for deformation of the optical path from the magnetostrictive strain of the galfenol layer.\\ 
 Despite the modest sensitivity this work achieves thermomechanically limited sensing with suspended waveguide coupling and a galfenol thin-film atop the optomechanical resonator whilst utilising portable electronics and DFB laser. 


\section{Conclusion}
%
%
Optomechanical magnetometers promise to enable a range of  research and industrial applications. Many of these will require fully integrated magnetometers operating with compact lasers and electronics. In this work we make progress towards this goal, demonstrating an optomechanical magnetometer that is integrated on a silicon chip with a suspended optical waveguide, utilises galfenol as a magnetostrictive material to provide improved resilience to corrosion and oxidation, and achieves 
thermomechanical noise-limited performance using a DFB laser and compact electronic systems. 
\begin{backmatter}
\bmsection{Funding}
The Commonwealth of Australia (represented by the Defence Science and Technology Group) supports this research through a Defence Science Partnerships agreement. This work was financially supported by the Australian Research Council (ARC) Centre of excellence for Engineered Quantum systems (EQUS): Grant No. CE170100009, and by Orica Australia \textit{Pty Ltd}.

\bmsection{Acknowledgments}
The Authors acknowledge the highly valuable advice and support provided by Rodney Appleby. The authors also  acknowledge the University of Queensland's Centre for Microscopy and Micro-analysis (CMM) and the Queensland node of the Australian National Fabrication Facility (ANFF). The equipment and staff expertise of the CMM and ANFF enabled the fabrication of the devices.

\bmsection{Disclosures}
The authors declare no conflicts of interest.


\end{backmatter}

\bibliography{References.bib}
\end{document}